\documentclass[prl,twocolumn,showpacs,amsmath,amssymb,superscriptaddress]{revtex4}%
\usepackage{graphicx}
\usepackage{dcolumn}
\usepackage{bm}
\newcommand{\rvec}{{\bf r}}
\newcommand{\qvec}{{\bf q}}
\newcommand{\Qvec}{{\bf Q}}
\newcommand{\kvec}{{\bf k}}
\newcommand{\beq}{\begin{equation}}
\newcommand{\eeq}{\end{equation}}
\newcommand{\beqa}{\begin{eqnarray}}
\newcommand{\eeqa}{\end{eqnarray}}

\begin{document}
\title{Odd parity charge density-wave scattering in cuprates}
\author{G. Seibold}
\affiliation{Institut f\"ur Physik, BTU Cottbus, PBox 101344, 03013 Cottbus, Ger
many}
\author{M. Grilli}
\affiliation{SMC-INFM-CNR and 
Dipartimento di Fisica, Universit\`a di Roma ``La Sapienza'', P.le Aldo
Moro 5, I-00185 Roma, Italy}
\author{J. Lorenzana}
\affiliation{SMC-INFM-CNR and 
Dipartimento di Fisica, Universit\`a di Roma ``La Sapienza'', P.le Aldo
Moro 5, I-00185 Roma, Italy}
\affiliation{ISC-CNR, Via dei Taurini 19, I-00185, Roma, Italy}
\date{\today}
\begin{abstract}
We investigate a model where superconducting electrons are coupled to a 
frequency dependent charge-density wave (CDW) order
parameter $\Delta_{\rvec}(\omega)$. Our approach can reconcile
the simultaneous existence of low energy Bogoljubov quasiparticles 
and high energy electronic order as observed in scanning tunneling 
microscopy (STM) experiments. The theory accounts for the
contrast reversal in the STM spectra between positive and
negative bias observed above the pairing gap. An intrinsic relation
between scattering rate and inhomogeneities follows
naturally. 
\end{abstract}

\maketitle
The phenomenon of charge ordering and its relation to superconductivity
remains a puzzling issue in the physics of high-$T_c$ cuprates.
Whereas the occurence of charge inhomogeneities was early on evidenced
by local probes and discussed in the context of electronic phase
separation\cite{erice} Tranquada and coworkers\cite{tra95}  
succeeded in probing a one-dimensional static spin and
lattice modulation in a rare earth codoped lanthanum
cuprate compound (LCO) by elastic neutron scattering. The unambigous
existence of an associated 
charge modulation has been established only more recently with 
resonant soft X-ray scattering\cite{abb05}.
In other cuprate compounds the existence of charge ordering 
so far comes from surface sensitive probes, like scanning tunneling
microscopy (STM) and angle-resolved photoemission spectroscopy (ARPES).
STM experiments performed on bismuthate and oxychloride superconductors see 
a complex modulation
of the local density of states (LDOS) both in the superconducting (SC)
state \cite{hoff02,how03,elroy05,hashi06,hana07,wise08} and above $T_c$ 
\cite{wise08,versh04,hana04}. In both cases one observes peaks in 
the Fourier transform of the 
real space LDOS at wave-vectors $Q=2\pi/(4a_0)...2\pi/(5a_0)$ suggestive of
checkerboard or stripe charge order. However, the debate is about the
question whether these peaks are non-dispersive in energy (and thus signature of
'real' charge order) or follow a bias-dependent dispersion due to quasiparticle
interference (QPI). In the latter case the spatial LDOS variations  
can be understood
from the so-called octet model \cite{hoff02,wang03}
which attributes the modulations to the 
elastic scattering between the high density regions of the 
Bogoljubov 'bananas' in the superconducting state.
Recent STM investigations \cite{kohsaka08,alldredge08} may resolve this apparent
conflict since they suggest that both, dispersive and non-dispersive
scattering originates from different regions in momentum and energy space.
The states in the nodal region 
which are well defined in k-space and undergo a transition
to a d-wave SC state below $T_c$ are then responsible for the low energy
QPI structure of the LDOS, whereas the ill-defined k-space 'quasiparticle'
states in the antinodal regions 
are responsible for the non-dispersive charge order above some energy 
scale $\Omega_0$. It is therefore a key issue to understand the nature
of this charge ordering, which seems particularly elusive, at least at low
energies, both in STM and ARPES experiments. 

In this paper we propose a phenomenological model which  
captures the above physical scenario by considering a frequency
dependent charge-density wave order parameter  analogous to the 
frequency dependent superconducting order parameter of Eliashberg
theory. By construction the order parameter  vanishes on the Fermi
surface so that for small energies the system appears homogeneous 
and the concept of  QPI applies while for large energies it appears to be 
charge ordered. 
As a bonus the system shows a strong high energy contrast reversal of
the LDOS as a function of energy in agreement with
experiments\cite{ma08} and an intimate relation between
inhomogeneities and quasiparticle scattering rate.  

We consider here a two-dimensional system of SC itinerant electrons
scattering with a charge density wave order parameter with an internal
dynamics. This is 
phenomenologically represented by a retarded local self-energy where the space
and frequency dependencies are factorized 
\beq
\Sigma_{\rvec}^{CDW}(\omega)=\Delta_{\rvec}(\omega)\equiv\Delta_{\rvec}^0+ v_{\rvec}^2 f(\omega).
\label{phenselfen}
\eeq

A conventional CDW without internal dynamics has $f(\omega)=0$ and 
$\Delta_{\rvec}^0\ne 0$. The function $f(\omega)$
describes the internal dynamics of the CDW, that is particle-hole
correlations building the charge density wave are time dependent.
The spatial dependence is implemented via the modulation of
$v_{\rvec}^2$. We will restrict to ${\rm Im} f(\omega)>0$, and
$v_{\rvec}$ real, which ensures ${\rm
  Im}\Sigma_{\rvec}^{CDW}>0$ as required for stability.  
The internal dynamics can be motivated on the basis of a frequency dependent
interaction, presumably of electronic origin and the same that
originates superconductivity but which we do not need to specify.
It implies that charge piling is {\em retarded} which
is physically appealing since it reduces the Coulomb penalty as in
Eliashberg theory of superconductivity. 
Such retardation effects are especially important for cuprates which are 
characterized by a large Coulomb repulsion.

In order to implement correctly the analytical properties of the
self-energy (Kramers-Kronig, etc.) it is convenient to make a pole
expansion,
$$f(\omega)=\sum_n\frac{1}{\omega-\epsilon^f_n-i\delta}.$$
This maps the self-energy to that of an effective 
Fano-Anderson model of itinerant electrons with conduction bandwidth
of order $t$ which can locally hop onto a distribution of $f$ levels via a
site-dependent hybridization term $v_{\rvec}$.
The bath of $f$ levels simulates the scattering due to the charge fluctuations
and provides a Hamiltonian formulation for the self-energy.

We will consider the dynamic case with $\Delta_{\rvec}^0=0$ and
compare with the conventional CDW model
($v_{\rvec}=0$, $\Delta_{\rvec}^0\ne 0$). A small $\Delta_{\rvec}^0$ 
component added to the frequency dependent case will not change our results
significantly. In addition we will take the modulation to be periodic 
$v_{\rvec}^2 =1/N_s \sum_n (v^2)_{n{\Qvec}} \exp(i n\Qvec \rvec)$.
Just as a superconductor with a frequency dependent order parameter
has long range off-diagonal order, even with $\Delta_{\rvec}^0=0$, the dynamical model
has diagonal order given by $n_\rvec=1/N_s \sum_{n,k} \int d\omega {\rm Im}
G_{\kvec,\kvec+n\Qvec}\cos(n\Qvec\rvec) $.  
In real materials the charge modulations will be linearly
coupled to the impurity potentials and the system will easily lose
long range order but keeping short range modulations with
translational symmetry breaking. This important effect can be 
incorporated but will be neglected for simplicity. 
 
Motivated by the experiments by  Kohsaka {\it et al.} \cite{kohsaka07} and previous
evidence on stripes\cite{tra95} we consider a one-dimensional
modulation with wave-vector ${|\Qvec|}=2\pi/4$ along the $x-$direction. 
Results for 
checkerboard patterns will be similar except for the absence of $C_4$ symmetry 
breaking.
The upper right insets to Fig. \ref{fig2}a,b  display
the associated modulation $\Delta^0_{\rvec}$ and $v^2_{\rvec}$. 
The resulting charge
modulation can be seen as bond-centered (hole) stripes separated by $4$ 
lattice constants
respectively. Our main conclusions do not depend on
this particular choice.

The Green function (GF) in k-space can be obtained from the
coupled system of equations

\begin{eqnarray}
&&\left[\omega - \varepsilon_{\kvec+n\Qvec} -v^2_0 f(\omega) 
\right]G_{\kvec+n\Qvec,\kvec+m\Qvec}^\sigma = \delta_{nm} \nonumber \\
&+&  \sum_{p\ne n}[\Delta^0_{(p-n)\Qvec}+  (v^2)_{(p-n)\Qvec} f(\omega)]G_{\kvec+p\Qvec,\kvec+m\Qvec}^\sigma \label{eq:2}
\end{eqnarray}
where the dispersion $\varepsilon_\kvec$ is measured with respect to
the chemical potential. While the zero momentum component of the 
static CDW order parameter $\Delta_{\qvec=0}^0$ gets trivially
reabsorbed in the chemical potential the same does not occur with the
dynamical part. That is 
$\Sigma_0^{CDW}(\omega) \equiv v^2_0 f(\omega)$, with  $v^2_0\equiv
(v^2)_{\qvec=0}$, has to be explicitely taken into
account as can be seen from the mapping to the hamiltonian structure 
which guarantees the preservation of sum rules for the
spectral function. Furthermore, as discussed above,  we are forced to
take $v^2_\rvec\ge0$ which implies that the Fourier component $v^2_0$
has to be positive for any non-zero modulation. 
Thus we have the surprising result that the dynamical CDW scattering
produces a momentum independent scattering rate.  
For a given amplitude of modulation, there is a lower bound for
such a scattering rate which is determined by taking the 
smallest $v^2_\rvec$ to be zero. This leads to a scattering rate
of the charge carriers by the CDW fluctuations $\Gamma_\omega^{CDW}\sim {\rm Im}
\Sigma^{CDW}_0(\omega)$. 
In the following we make this minimum choice which makes
$\Gamma_\omega^{CDW}$ to be explicitely determined by the sum of
amplitudes of the electronic inhomogeneity $\sim v^2_0$. 
Such an intrinsic relation between electronic inhomogeneity and
inelastic scattering rate has been recently revealed by STM experiments
on Bi2212 materials \cite{alldredge08} where it has been shown 
that the LDOS spectra can be parametrized based on
a model with SC d-wave order supplemented by  
an energy dependent scattering rate $\Gamma^{LDOS}_\omega=\alpha \omega$.
The parameter $\alpha$ varies spatially and in the regions with
pronounced charge order acquires values up to $\alpha \approx 0.4$.
Also  ARPES spectra\cite{valla99,johnson01} from Bi2212 materials
suggest a  marginal Fermi liquid (MFL) type self-energy
(i.e. ${\rm Im} \Sigma\sim \Gamma_\omega^{ARPES} \sim \omega$) which is increasing 
with underdoping similar to $\Gamma^{LDOS}_\omega$.
Motivated by these experimental findings we describe the CDW dynamics
(and therefore $\Gamma_\omega^{CDW}$) by a marginal Fermi liquid (MFL) type 
structure \cite{varma91} 
\begin{equation}\label{eq:mfl}
f(\omega)= 2\omega \ln\frac{\Gamma+i\omega}{\Omega} +i \pi \Gamma
\end{equation}
where $\Gamma=kT$ corresponds to a temperature scale and $\Omega$ 
denotes an upper cutoff of the boson spectrum from which the MFL
self-energy is derived\cite{note}.
The amplitudes of the charge modulation are chosen such that
$v^2_0= 0.08 eV$ which yields ${\rm Im} \Sigma(\omega)=v^2_0
{\rm Im}f(\omega)\approx
0.25 |\omega|$ at small frequencies which is close to the average 
$\Gamma^{LDOS}_\omega$ observed. 
The dispersion $\varepsilon_\kvec$ includes hopping to nearest 
($\sim t_1=250$meV) and next nearest ($t_2=-0.4t_1$) neighbors
as appropriate for Bi2212 materials. 

\begin{figure}[bt]
\includegraphics[width=8cm,clip=true]{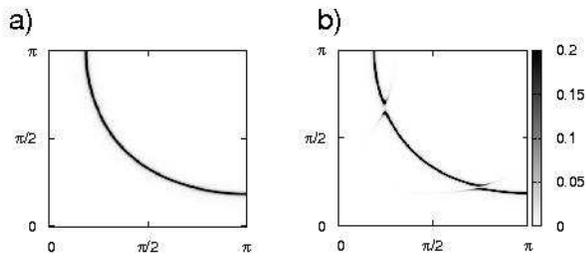}
\caption{Density plots of the 
integrated spectral weight in 
a window of $\pm 5 $meV around $E_F$. (a) dynamic CDW ($\Delta_{i}^0=0$);
(b) static CDW with $\Delta^0_{2}-\Delta^0_{1}=0.054 eV$ (cf. upper
right inset to Fig. \ref{fig2}). Residual parameters as given in the text.
}
\label{fig1}
\end{figure}

Fig.~\ref{fig1}a demonstrates that for the dynamic case 
quasiparticles around the Fermi energy are protected from the CDW scattering. 
In contrast, static CDW order (cf. Fig.~\ref{fig1}b) produces
the usual FS reconstruction, i.e. flattening of the dispersion 
around $(\pi,0)$ and gap formation 
between the nodes and $(0,\pi)$ due to nesting. At higher
energies  the effect of the dynamic scattering on the spectral
function is similar to 
that of a static CDW, with an additional
broadening due to the imaginary
part of the self-energy which is increasing with frequency.

\begin{figure}[tb]
\includegraphics[width=7cm,clip=true]{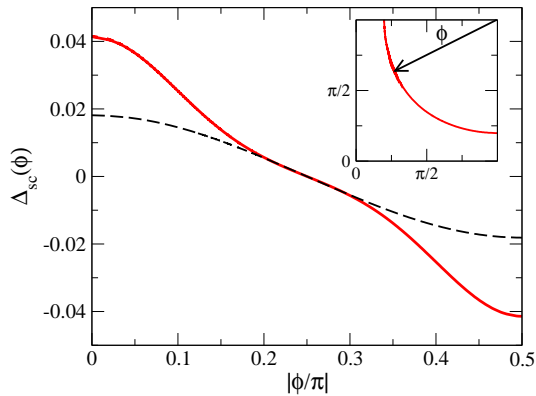}
\caption{(Color online) Full line: Anharmonic SC gap structure 
Eq. (\protect\ref{scgap}) implemented
in the calculation of the LDOS; Dashed: Pure d-wave contribution to
the anharmonic gap around the nodes ($\Phi/\pi=0.25$).}
\label{fig1b}
\end{figure}

\begin{figure}[tb]
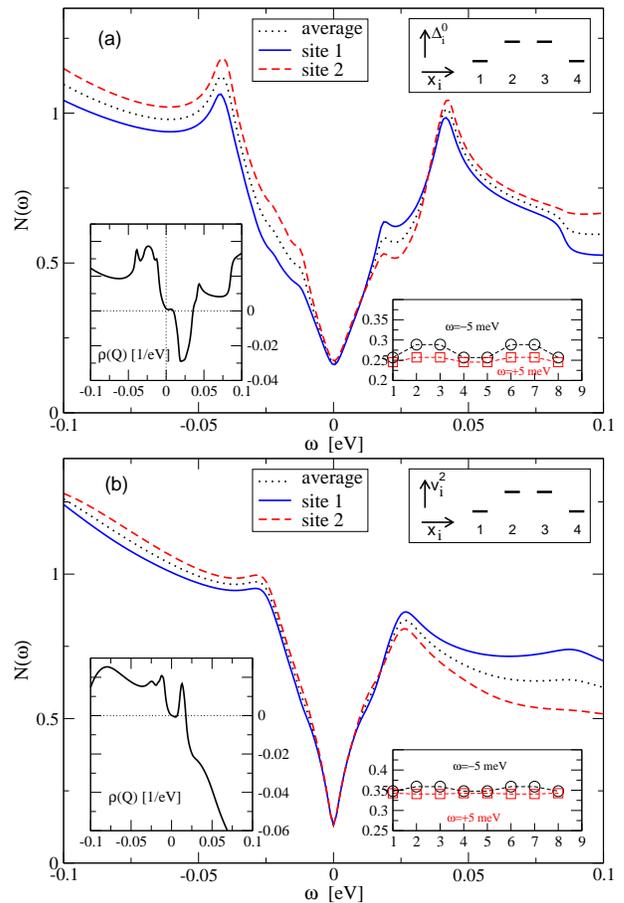

\includegraphics[width=8cm,clip=true]{fig3a.eps}
\includegraphics[width=8cm,clip=true]{fig3b.eps}
\caption{(Color online) LDOS for the model with static 
CDW scattering $\Delta^0_{2}-\Delta^0_{1}=0.054 eV$ (a) 
and  frequency dependent CDW scattering $f(\omega)$ (b). 
Upper insets: Sketch of the underlying modulation of $v^2_\rvec$.
Lower left insets: Fourier tranformed of the LDOS at
the CDW scattering vector $\Qvec=\frac{2\pi}{4}$. 
The phase has been chosen
such that ${\rm Im} \rho(\Qvec)=0$.
Lower right insets: LDOS at $\omega=+5$meV (squares) and
$\omega=-5$meV (circles). 
Further parameters:
Chemical potential $\mu=-0.23 eV$ 
(doping $x\approx 0.07$), $\Gamma=1meV$ and $\Omega=1eV$. 
}
\label{fig2}
\end{figure}

As mentioned above the interpretation of STM experiments, which are
usually taken at very low temperatures, requires the implementation
of superconductivity  into the formalism which in Eq. \ref{eq:2} can be
added as a BCS-type SC self-energy 
$\Sigma_{\kvec}^{sc}=(\Delta^{sc}_k)^2/[\omega+
\varepsilon_\kvec+v_0^2f^*(-\omega)]$ to the bracket on the l.h.s.
Recent STM experiments \cite{pulp09} have revealed that the
shape of the tunneling spectra of Bi2212 can be fit by including higher
harmonic contributions in addition to the simple d-wave form.
Following this observation  we take 
\begin{equation}\label{scgap}
\Delta^{sc}_k=\sum_n \Delta_0(n) \left[\cos(nk_x) -\cos(nk_y)\right]/2
\end{equation}
and restrict to the first three harmonics with
$\Delta_0(1)=40 meV$, $\Delta_0(2)=-10 meV$, and $\Delta_0(3)=5
meV$. For simplicity we take this gap function to be frequency
independent. 
The resulting gap structure is shown in Fig.~\ref{fig1b} in comparison
with a pure d-wave form ($\Delta_0(1)=18 meV$) which fits the
anharmonic gap in the nodal region. The larger gap in the antinodes is
supposed to be due to local pairing and persist above $T_c$ producing
the pseudogap.

In Fig.~\ref{fig2} we report the LDOS structure for the model
with static (a) and dynamic (b) order parameter including
the SC gap. 
The static order parameter was chosen in order to give a similar
magnitude of the intensity difference among different sites
(contrast) at high positive bias as the dynamical model.  

Due to particle hole mixing superconductivity tends to soft the charge
order related LDOS modulations at low energies, however the inset in
(a) reveals that in the static case 
substantial residual modulations persist at an energy
of $\omega=\pm 5 $meV specially for negative energies.
This contrast is strongly
suppressed for the dynamic order parameter [see inset in (b)] as seen
experimentally. This is easy to understand due to the structure of
$f(\omega)$ which vanishes at low energies. 
However, this effect is not very apparent and would hardly allow
one to decide for the dynamical model. 
A much more dramatic difference is instead provided by the sign of the
contrast when the bias is reversed. For the static order parameter we
observe contrast reversal below the energy of the superconducting gap
(i.e. sites with high intensity and low intensity LDOS are inverted
when one changes the sign of the bias) while the 
ordering of the intensities tends to be substantially preserved  
at high energies (contrast preservation). For the dynamical case the
situation is the opposite: there one finds contrast preservation inside 
and contrast reversal above the superconducting
gap. Thus the dynamical model agrees with the ubiquitous contrast
reversal found in STM at large energy concomitant with the absence 
at low energies\cite{ma08}.

How general is this result?
In the static case contrast reversal at high energy can be
achieved for special situations (e.g. half-filled system with square Fermi surface, i.e.
complete nesting) where asymmetry occurs for higher energies.
In these cases the imaginary part of the off-diagonal GF ${\rm Im} G_{k,k+Q}$ has a sign
change at the chemical potential since $k$-states with
$\varepsilon_k-\mu<0$ only connect to $k+Q$-states above the 
chemical potential and vice versa. 
In case of the dynamical order parameter
the sign change has its origin in the dynamical self-energy since
$G_{k,k+Q} \sim f(\omega)$ 
and ${\rm Re} f(\omega)=- {\rm Re} f(-\omega)$ and thus, it does not require special conditions as
nesting. Indeed  we have found similar results for other periodicities and
fillings.

The hump feature on the positive frequency side of the
LDOS in Fig.~\ref{fig2} originates from the nesting of 
$\Qvec$ between 
the antinodal FS segments. This induces a concomitant increase of the LDOS 
modulation.which for our underdoped system occurs at energies
$\approx 90$ meV above $E_F$.

To analyze the difference between frequency dependent and frequency
independent CDW scattering 
in more detail we show in the lower left insets to Figs. \ref{fig2} 
the Fourier transformed LDOS at multiples at the CDW wave vector.

As pointed out in Refs.~\cite{podo03,chen04} the symmetry of the wave-function
in a SC with static CDW order implies 
$\rho(\Qvec,\omega)=-\rho(\Qvec,-\omega)$ for sufficiently small $\omega$.
In fact, we clearly observe this asymmetry in the LDOS in Fig.~\ref{fig2}(a)
(black, solid curve in the lower left inset)
on the energy scale of the SC gap in contrast to the experimental
observation\cite{how03} of a symmetric
$\rho(\Qvec,\omega)$ in this regime.  This was explained as due to a
modulation of the SC pair density\cite{chen04,mach02}. Fig.~\ref{fig2}(b)
shows that the symmetry is also achieved for a dynamic CDW order
parameter. Notice that this symmetry can not persist up to high
energies since contrast reversal  implies that at high energy the dominant
Fourier component should change sign. Indeed our dynamic model with  
$\Delta^{CDW}(\omega) \sim f(\omega)$ with 
${\rm Re} f(\omega)=-{\rm Re}f(-\omega)$ reconciles both aspects, low energy symmetry
of  $\rho(\Qvec,\omega)$ and a strong high energy odd component leading to
contrast reversal.

The present model applies to cuprates where the system at the
microscopic scale shows breaking of translational symmetry but 
sustains low energy quasiparticles weakly affected by that.  
A prominent example is La$_{1.875}$Ba$_{0.125}$CuO$_4$ where a
surprisingly well defined Fermi surface has recently been
observed\cite{he09}.  In addition, there are situations with no apparent  
translational symmetry breaking but with a 
 high energy response resembling that of an ordered system. We have recently
presented a scenario for the spectral function in this distinct physical
situation\cite{gri09}.

  Summarising we have presented a simple phenomenological model of a
  frequency dependent CDW order parameter in cuprates which accounts
  for the dichotomy between low energy Bogoljubov quasiparticles and
  high energy electronically  ordered states \cite{kohsaka08} with the
  concomitant crossover from LDOS modulations determined from QPI to
  dynamical CDW scattering at large frequencies. For simplicity we have
  taken into account a charge order parameter but we expect similar results will
  apply for a magnetic order parameter.  The model has a
  series of implications which  agree with experimental observation:
  {\it i}) Intimate relation between quasiparticle scattering rate and
  amplitude of charge order.  {\it ii}) Suppression of contrast in the LDOS
  at low energies. {\it iii}) Contrast reversal (preservation) at high (low)
  energies. {\it iv}) Even symmetry of the low energy Fourier transformed
  LDOS. {\it v}) Featureless Fermi surface.  While a naive CDW order
  parameter will be in contradiction with most of these observations
  the dynamical model reconciles the long suspected charge ordering
  with these apparently contrasting experiments.  Our results for the
  scattering time suggest to search for the origin of marginal Fermi
  liquid behavior in an Eliashberg treatment of charge (and) or spin density
  waves\cite{guber93}.

This work has been supported by MIUR, PRIN 2007 (prot. 2007FW3MJX003)
and partially by NSF grant PHY05-51164 at KITP. 
J.L  thanks KITP-UCSB for hospitality under the program ``The Physics
of Higher Temperature Superconductivity''. M.G and G. S. 
acknowledge support from the Vigoni foundation.


\begin{thebibliography}{10}
\bibitem{erice} Proceedings of the {\it Third workshop on phase 
              separation, electronic inhomogeneitites and related mechanisms
              in high-$T_c$ superconductors}, C. Di Castro and E. Sigmund
              (eds.); J. Supercond. {\bf 9} (1996).
\bibitem{tra95}
J.~M. Tranquada, {\it et al.}, Nature (London) {\bf 375},  561  (1995).
\bibitem{abb05}
P. Abbamonte,  {\it et al.}, Nature Phys. {\bf 1},  155  (2005).
\bibitem{hoff02} J. E. Hoffmann,  {\it et al.}, Science {\bf 297}, 1148 (2002).
\bibitem{how03} C. Howald,  {\it et al.}, Phys. Rev. B {\bf 67}, 014533 (2003).
\bibitem{elroy05} K. McElroy,  {\it et al.}, Phys. Rev. Lett. {\bf 94}, 197005 (2005).
\bibitem{hashi06} A. Hashimoto,  {\it et al.}, Phys. Rev. B {\bf 74}, 064508 (2006).
\bibitem{hana07} T. Hanaguri,  {\it et al.}, Nature Physics {\bf 3}, 865 (2007).
\bibitem{wise08} W. D. Wise,  {\it et al.}, Nature Physics {\bf 4}, 696 (2008).
\bibitem{versh04} M. Vershinin,  {\it et al.}, Science {\bf 303}, 1995 (2004).
\bibitem{hana04} T. Hanaguri,  {\it et al.}, Nature {\bf 430}, 1001 (2004).
\bibitem{wang03} Q. H. Wang and D.-H. Lee, Phys. Rev. B {\bf 67}, 020511(R) (2003).
\bibitem{kohsaka08} Y. Kohsaka,  {\it et al.}, Nature {\bf 454}, 1072 (2008).
\bibitem{alldredge08} J. W. Alldredge,  {\it et al.}, Nature Physics {\bf 4}, 319 (2008).
\bibitem{kohsaka07} Y. Kohsaka,  {\it et al.}, Science {\bf 315}, 1380 (2007).
\bibitem{valla99} T. Valla,  {\it et al.}, Science {\bf 285}, 2110 (1999).
\bibitem{johnson01} P. D. Johnson,  {\it et al.},
  Phys. Rev. Lett. {\bf 87}, 177007 (2001).
\bibitem{note} Notice that the definitions of  $\Gamma^{LDOS}_\omega$,  
$\Gamma^{ARPES}_\omega$ and $\Gamma^{CDW}_\omega$ are not the same but
are closely related. For example introducing a MFL form in
$\Gamma^{CDW}_\omega$ automatically generates a MFL in $\Gamma^{ARPES}_\omega$.
 Eq, \ref{eq:mfl} only holds close to the chemical potential
whereas the calculation of e.g. particle numbers requires an
appropriate cutoff at higher energies.  
\bibitem{varma91} P. B. Littlewood and C. M. Varma, J. Appl. Phys. {\bf 69}, 4979 (1991).
\bibitem{pulp09} A. Pushp {\it et al.}, Science {\bf 324}, 1689 (2009).
\bibitem{podo03} D. Podolsky, E. Demlder, K. Damle, and B. I. Halperin, Phys. Rev. B {\bf 67}, 094514 (2003).
\bibitem{chen04} H.-D. Chen, O. Vafek, A. Yazdani, and S.-C. Zhang, Phys. Rev. Lett. {\bf 93}, 187002 (2004).
\bibitem{mach02} M. Ichioka and K. Machida,J. Phys. Soc. Jpn. {\bf 71}, 1836 (2002).
\bibitem{ma08} J.-H. Ma,  {\it et al.}, Phys. Rev. Lett. {\bf 101}, 207002 (2008).
\bibitem{he09}R.-H. He {\it et al.}, Nat. Phys. {\bf 5}, 119 (2009).  
\bibitem{gri09} M. Grilli, G. Seibold, A. Di Ciolo, and J. Lorenzana,
Phys. Rev. B {\bf 79}, 125111 (2009).

\bibitem{guber93} P. Niyaz, J. E. Gubernatis, R. T. Scalettar, and C. Y. Fong,
Phys. Rev. B{\bf 48}, 16011 (1993).


\end{thebibliography}
\end{document}